\documentclass{article}
\usepackage{spconf,amsmath,graphicx}
\usepackage{multirow}
\usepackage{amssymb}
\usepackage{xcolor}
\usepackage[hidelinks]{hyperref}

\usepackage{enumitem}
\setlist{nosep, leftmargin=14pt}

\usepackage{mwe} 

\usepackage[numbers,sort&compress]{natbib}   

\title{Bonnet: Ultra-Fast Whole-Body Bone Segmentation from CT Scans}
\name{%
\begin{tabular}{c}
Hanjiang Zhu$^{1,2,3}$, Pedro Martelleto Rezende$^{4}$, Zhang Yang$^{5,6}$, Tong Ye$^{3}$\\
Bruce Z. Gao$^{3}$, Feng Luo$^{3}$, Siyu Huang$^{3}$, Jiancheng Yang$^{1,2}$%
\end{tabular}%
\thanks{Corresponding author: Siyu Huang.}%
\thanks{This work was conducted during Hanjiang Zhu's research internship.}%
}

\address{
$^1$ ELLIS Institute Finland, Finland~~~~~~
$^2$ Aalto University, Finland\\
$^3$ Clemson University, USA~~~~~~ 
$^4$ ETH Zurich, Switzerland\\
$^5$ Fujian Medical University Union Hospital, China~~~~~~ 
$^6$ Fujian Medical University, China 
} 
\begin{document}
\maketitle
\begin{abstract}
This work proposes Bonnet, an ultra-fast sparse-volume pipeline for whole-body bone segmentation from CT scans. Accurate bone segmentation is important for surgical planning and anatomical analysis, but existing 3D voxel-based models such as nnU-Net and STU-Net require heavy computation and often take several minutes per scan, which limits time-critical use. The proposed Bonnet addresses this by integrating a series of novel framework components including HU-based bone thresholding, patch-wise inference with a sparse spconv-based U-Net, and multi-window fusion into a full-volume prediction. Trained on TotalSegmentator and evaluated without additional tuning on RibSeg, CT\text{-}Pelvic1K, and CT\text{-}Spine1K, Bonnet achieves high Dice across ribs, pelvis, and spine while running in only 2.69 seconds per scan on an RTX~A6000. Compared to strong voxel baselines, Bonnet attains a similar accuracy but reduces inference time by roughly $25\times$ on the same hardware and tiling setup. 
The toolkit and pre-trained models will be released at \texttt{https://github.com/HINTLab/Bonnet}.
\end{abstract}

\begin{keywords}
Bone segmentation, CT scan, computed tomography, sparse convolution, fast inference.
\end{keywords}
\section{Introduction}
\label{sec:intro}
Accurate segmentation of skeletal structures from CT images is vital for medical analysis and clinical applications, as bone structures inform surgical planning, postoperative reconstruction, and pathological assessment. They also serve as a geometric foundation for anatomical landmark localization, rigid-body registration, and biomechanical modeling. Owing to their high density and geometric stability in CT imaging, bones are often used as structural references for multi-organ modeling and spatial alignment~\cite{minnema2022review,kuiper2023osteotomy,inproceedings}.
\begin{figure}[t]
  \centering
  \includegraphics[width=\linewidth]{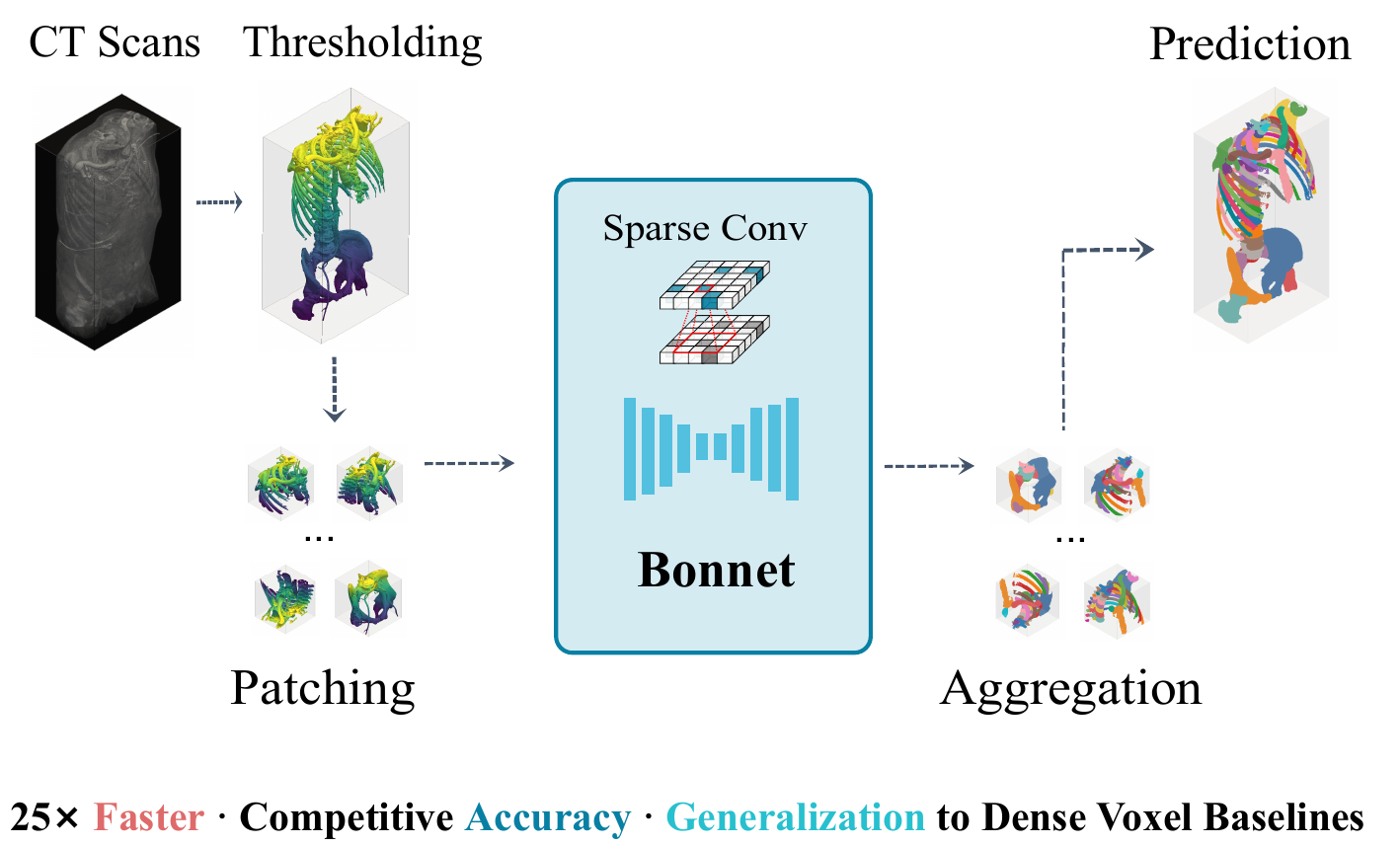}
  \caption{Overview of the Bonnet pipeline. Bonnet first applies simple HU thresholding to obtain a sparse CT volume, then segments bone structures patch-wise using a U-Net backbone built with sparse 3D convolutions, and finally aggregates the patch predictions into the final segmentation result. Bonnet delivers \(\sim\!25\times\) faster inference while maintaining anatomical accuracy and robust generalization.}
  \label{fig:model}
\end{figure}
However, achieving high-precision whole-body bone segmentation remains challenging. On the one hand, the human skeleton contains a large number of bones with highly diverse morphologies, making it difficult for traditional 2D convolutional or patch-based 3D convolutional networks to simultaneously capture global structural consistency and local detail accuracy. On the other hand, dense voxel-based convolutional models (such as 3D U-Net~\cite{9018080} and its variants) incur enormous computational costs and memory usage when processing high-resolution whole-body CT volumes, resulting in minute-scale inference times that hinder real-time clinical applications.

To address these limitations, recent studies have explored sparse voxel or point-cloud representations that exploit the inherent sparsity of bone tissues in CT volumes. Among them, RibSeg~\cite{ribsegv1} and its extension RibSeg v2~\cite{ribsegv2} established pioneering large-scale benchmarks for rib segmentation and labeling. These works demonstrated that point-cloud-based models~\cite{qi2017pointnetdeephierarchicalfeature,yang2019modeling} can achieve comparable accuracy to dense voxel networks while offering faster inference. Although these studies verified the feasibility of point-based learning for bony structures, they were limited to a single anatomical region (ribs) and did not explore scalability or generalization across the entire skeletal system.

Following these advances, several studies extended the point-cloud paradigm to other anatomical structures.
Zhu et al.~\cite{s24134076} introduced an adaptive denoising and smoothing method for rib reconstruction based on CT-derived point clouds.
Zhou et al.~\cite{Zhou2025PointCloud} applied point-cloud segmentation to pelvic organ delineation, achieving more than a tenfold acceleration over voxel-based baselines while maintaining high accuracy.
Xie et al.~\cite{XIE2025103367} combined point-cloud, graph, and implicit-field representations to develop an efficient segmentation framework for pulmonary airway labeling.
These works collectively highlight the potential of sparse and point-cloud representations in medical imaging, yet they remain restricted to localized anatomical regions and lack a comprehensive framework for multi-scale, cross-domain skeletal segmentation.

To overcome these challenges, we introduce Bonnet, a fast sparse-volume inference pipeline for whole-body bone segmentation. As shown in Fig.~\ref{fig:model}, Bonnet turns the CT into a sparse volume using HU-based bone thresholding and performs patch-wise sparse inference with a lightweight spconv U-Net~\cite{graham2017submanifoldsparseconvolutionalnetworks,graham20173dsemanticsegmentationsubmanifold}. This yields whole-body bone masks in only 2.69 seconds per scan on a single RTX~A6000, whereas strong voxel baselines typically require more than one minute per scan.

Because Bonnet is both fast and robust across anatomically distinct skeletal regions, it can still deliver usable bone segmentation in time-critical or suboptimal settings (e.g., trauma triage CT, intraoperative guidance, emergency planning, large-scale retrospective analysis). In such scenarios, waiting $>60$\,s for a full 3D voxel model is often impractical, whereas a 2--3\,s result is clinically usable.

We train Bonnet \emph{from scratch} on a single public whole-body dataset (the TotalSegmentator~\cite{doi:10.1148/ryai.230024} training split) and then directly apply it, \emph{without any retraining or fine-tuning}, to independent external CT bone datasets that focus on different anatomical regions, including RibSeg~\cite{ribsegv2} for ribs, CT\text{-}Pelvic1K~\cite{liu2021deeplearningsegmentpelvic} for the pelvis, and CT\text{-}Spine1K~\cite{deng2024ctspine1klargescaledatasetspinal} for the spine. As shown later in Sec.~\ref{sec:exp}, Bonnet maintains high Dice under this internal$\rightarrow$external shift while keeping inference in the single-digit second range.

We summarize our contributions as follows.
\begin{itemize}
    \item We revisit dense voxel convolutions through the lens of sparsity, and propose Bonnet, a fast sparse-volume inference pipeline that achieves \(\sim\!25\times\) faster inference while maintaining high accuracy;
    \item We systematically validate the method's adaptability and cross-domain generalization across multiple anatomical regions and datasets;
    \item We implement a lightweight and robust sparse bone toolkit that enables second-scale inference, providing an efficient and scalable foundation for large-scale anatomical analysis and downstream medical imaging applications. It will be released as an open-source toolkit for CT bone segmentation.
\end{itemize}

\section{Method}
\label{sec:methods}
\subsection{Overview}
This work presents Bonnet, which performs whole-body bone segmentation in three stages: (i) HU-based bone thresholding to obtain a sparse CT volume; (ii) patch-wise sparse inference using a spconv-based U-Net backbone followed by a lightweight segmentation head; and (iii) aggregation of per-patch predictions into a full-volume output. The rest of this section details the sparse sliding-window inference, multi-window fusion, and sparse convolutional architecture.

\subsection{Sparse Voxel Sliding-Window and Fusion}

Guided by a bone HU prior, we keep only voxels with HU in $[200,3000]$. Voxel intensities are standardized with training-set statistics $(\mu,\sigma)$ via z-score to form point-wise features. Each case is cached as a sparse triplet \emph{(coordinates, features, labels)} for reproducibility and fast loading.

During training, we randomly crop sparse windows of size $128^3$: with probability $1-\rho$ (we set $\rho{=}0.33$) we sample uniformly; with probability $\rho$ we prioritize windows near bone foreground to better cover thin/small structures. At test time, we use overlapping sliding windows with overlap $0.5$, run forward per window, and fuse globally.

For a voxel $\mathbf{x}$, let $p_t(c\!\mid\!\mathbf{x})$ be the softmax score from the $t$-th window. We fuse scores by
\begin{equation}
\hat{p}\!\left(c\mid \mathbf{x}\right)
=\sum_{t:\,\mathbf{x}\in\mathcal{S}_{\mathbf{w}}^{(t)}}
a_t(\mathbf{x})\,p_t\!\left(c\mid \mathbf{x}\right).
\end{equation}
The final prediction is
\begin{equation}
\hat{y}(\mathbf{x})=\arg\max_{c}\,\hat{p}\!\left(c\mid \mathbf{x}\right).
\end{equation}
To downweight boundary uncertainty and repeated coverage, we use a Gaussian decay
\begin{equation}
a_t(\mathbf{x})
=\exp\!\left(
-\frac{\left\lVert \mathbf{x}-\mathbf{m}_t\right\rVert_2^2}
     {2\,(\sigma\cdot \mathbf{w})^2}
\right),
\end{equation}
where $\mathbf{m}_t$ is the window center, $\mathbf{w}$ is the window edge-length vector, and we set the Gaussian decay parameter $\sigma$ to 0.5 in all experiments.

\subsection{Sparse U-Net Architecture and Sparse Convolutions}
We adopt a spconv-based sparse U-Net. The encoder alternates Submanifold Sparse Convolutions (SubMConv) and stride-2 Sparse Convolutions for feature extraction and downsampling; the decoder uses Inverse Sparse Convolutions for upsampling and symmetric skip connections. Each block is followed by \emph{SparseInstanceNorm} and \emph{LeakyReLU(0.01)}. Channel widths are determined by a width factor of $4.0$. The top sparse features are fed into a lightweight MLP head to produce $K$-class logits. Inputs contain only thresholded bone voxels and are standardized by $(\mu,\sigma)$.

Let $\mathcal{S}\subset\Omega$ be the sparse support and $\{(\mathbf{x}_i,\mathbf{h}_i)\}_{i=1}^{N}$ the feature set. Given a kernel $\mathcal{K}\subset\mathbb{Z}^3$, a sparse convolution is
\begin{equation}
\mathbf{h}'(\mathbf{x})=\phi\!\left(
\sum_{\boldsymbol{\delta}\in\mathcal{K}}
W_{\boldsymbol{\delta}}\;\mathbf{h}(\mathbf{x}+\boldsymbol{\delta})\;
\mathbf{1}_{\mathcal{S}}(\mathbf{x}+\boldsymbol{\delta})
\right),\quad \mathbf{x}\in\mathcal{S},
\end{equation}
where $\phi$ denotes normalization and activation. SubMConv updates only on the original support $\mathcal{S}$; stride-2 Sparse Conv builds coarser supports (downsampling); Inverse Sparse Convolutions map features back to finer grids and concatenate with encoder features at the same scale.

\subsection{Learning Objective}
We combine cross-entropy with label smoothing and Soft Dice. Let the number of classes be $K$, sample index $i$, ground-truth label $y_i\in\{0,\dots,K-1\}$, and softmax probability $p_i(c)$.
\noindent{Cross-entropy with label smoothing is:}
\begin{equation}
\mathcal{L}_{\mathrm{CE}}
= -\sum_{i}\sum_{c=0}^{K-1} q_i(c)\,\log p_i(c),
\end{equation}
\begin{equation}
q_i(c)=
\begin{cases}
1-\varepsilon, & c=y_i,\\[4pt]
\dfrac{\varepsilon}{K-1}, & c\neq y_i,
\end{cases}
\end{equation}
\noindent{The Soft Dice metric is:}
\begin{equation}
\mathrm{SoftDice}
= \frac{1}{K-1}\sum_{c=1}^{K-1}
\frac{2\sum_i p_i(c)\,\mathbf{1}[y_i=c]+\epsilon}{
\sum_i p_i(c)+\sum_i \mathbf{1}[y_i=c]+\epsilon},
\end{equation}
where $\epsilon$ ensures numerical stability. In practice, we sum $\mathcal{L}_{\mathrm{CE}}$ and $(1-\mathrm{SoftDice})$ with equal weights for each sparse window.

\section{Experiments}
\label{sec:exp}

\subsection{Datasets}
We train Bonnet on the TotalSegmentator dataset, using a fixed split of 911/228/89 scans for training/validation/testing. TotalSegmentator provides full-body CT scans with comprehensive multi-organ labels, including detailed bone annotations. 

To assess cross-region and cross-domain generalization without additional fine-tuning, we directly evaluate the trained model on three independent external datasets: RibSeg (160 CT cases focusing on rib-level rib cage parsing), CT\text{-}Pelvic1K (a pelvic CT set with clinically relevant bone structures, we use the 103-case clinical subset), and CT\text{-}Spine1K (a large-scale spine CT benchmark with vertebral labels, evaluated on its 197-case official test split). These datasets target anatomically distinct skeletal regions (ribs, pelvis, spine), allowing us to test whether Bonnet can generalize beyond the distribution it was trained on.
\begin{table}[!t]
\centering
\caption{Dice (\%) on the TotalSegmentator test set (89 scans). Columns report Dice for rib (24), pelvis (3), spine (25), and all bones (62). \textit{Forward (s)} is the per-scan inference time on an RTX~A6000.}
\label{tab:indomain}
\setlength{\tabcolsep}{3pt}
\resizebox{\linewidth}{!}{%
\begin{tabular}{lccccc}
\hline
\textbf{Model} & \textbf{Rib (24)} & \textbf{Pelvis (3)} & \textbf{Spine (25)} & \textbf{Overall (62)} & \textbf{Forward (s)} \\
\hline
nnU-Net             & 89.4 & 97.82 & 90.12 & 94.54 & 66.87 \\
nnU-Net (RAW)      & 90.45 & 97.68 & 90.45 & 94.13 & 92.83 \\
STU-Net-S         & 86.97 & 96.88 & 86.68 & 92.72 & 71.32 \\
STU-Net-B         & 88.37 & 97.56 & 90.18 & 94.31 & 74.92 \\
STU-Net-L         & 91.54 & 97.96 & 89.21 & 94.94 & 77.72 \\
STU-Net-S-ft      & 90.67 & 96.84 & 87.42 & 93.58 & 71.45 \\
STU-Net-B-ft      & 91.53 & 97.55 & 89.74 & 94.41 & 74.86 \\
STU-Net-L-ft      & 91.91 & 98.78 & 91.28 & \textbf{95.52} & 78.12 \\
\hline 
PointNet          & 56.57 & 95.44 & 52.63 & 75.24 & 4.51 \\
PVCNN             & 79.5 & 98.82 & 81.66 & 89.93 & 8.83 \\
SPVCNN            & 75.35 & 96.27 & 77.76 & 85.39 & \textbf{1.05} \\ \hline 
\textbf{Bonnet (Ours)}& 94.25 & 99.64 & 94.32 & 94.91 & 2.69 \\
\hline
\end{tabular}%
}
\end{table}
\subsection{Experimental Setup}
All models are trained on a single NVIDIA RTX~A6000 GPU. Preprocessing keeps only voxels with HU in $[200,3000]$ and standardizes intensities using the training-set mean and standard deviation. Inference uses sparse $128^3$ sliding windows with $0.5$ overlap and Gaussian-decay fusion (decay factor $0.5$).

We evaluate Bonnet for segmenting whole-body bone structures spanning 62 labeled components, including: \textbf{Rib} (24 individual rib segments across the rib cage); \textbf{Spine} (25 vertebral elements along the spinal column); \textbf{Pelvis} (3 pelvic structures: left hip, right hip, and sacrum); \textbf{Femur} (2 femora: left and right femur); \textbf{Humerus} (2 humeri: left and right humerus); \textbf{Scapula} (2 scapulae: left and right scapula); \textbf{Clavicle} (2 clavicles: left and right clavicle); as well as the \textbf{Skull} and the \textbf{Sternum}. We report Dice~(\%) for anatomically meaningful groups such as Rib, Spine, Pelvis, and also for the union of all these labeled structures, denoted as \textbf{Overall (62 classes)} in Tables~\ref{tab:indomain} and \ref{tab:ood}. Runtime (\textbf{Forward (s)}) is measured as the forward-pass inference time per scan.

\begin{table}[!t]
\centering
\caption{Cross-dataset Dice (\%). ``Internal'' denotes performance on the TotalSegmentator test set. ``External'' evaluates the same models on an out-of-domain dataset for each region: ribs on RibSeg, pelvis on CT\text{-}Pelvic1K, and spine on CT\text{-}Spine1K.}
\label{tab:ood}
\setlength{\tabcolsep}{3pt}
\resizebox{1\linewidth}{!}{%
\begin{tabular}{lcccccc}
\hline
\multirow{2}{*}{\textbf{Model}} & \multicolumn{2}{c}{\textbf{Rib (24)}} & \multicolumn{2}{c}{\textbf{Pelvis (3)}} & \multicolumn{2}{c}{\textbf{Spine (25)}} \\
 & Internal & External & Internal & External & Internal & External \\
\cline{1-7}
nnU-Net            & 89.4 & 86.21 & 97.82 & 96.6 & 90.12 & 90.06 \\
nnU-Net (RAW)      & 90.45 & 85.97 & 97.68 & 96.63 & 90.45 & 89.84 \\
STU-Net-S         & 86.97 & 85.47 & 96.88 & 96.25 & 86.68 & 85.97 \\
STU-Net-B         & 88.37 & 86.06 & 97.56 & 96.41 & 90.18 & 88.52 \\
STU-Net-L         & 91.54 & 86.32 & 97.96 & 96.44 & 89.21 & 90.11 \\
STU-Net-S-ft      & 90.67 & 85.76 & 96.84 & 96.54 & 87.42 & 86.44 \\
STU-Net-B-ft      & 91.53 & 86.09 & 97.55 & 96.63 & 89.74 & 88.99 \\
STU-Net-L-ft      & 91.91 & \textbf{86.41} & 98.78 & 96.7 & 91.28 & 90.17 \\
\cline{1-7}
PointNet          & 56.57 & 6.68 & 95.44 & 30.02 & 52.63 & 10.19 \\
PVCNN             & 79.5 & 2.63 & 98.82 & 2.35 & 81.66 & 1.24 \\
SPVCNN            & 75.35 & 39.38 & 96.27 & 70.29 & 77.76 & 25.23 \\ \hline 
\textbf{Bonnet (Ours)} & 94.25 & 85.91 & 99.64 & \textbf{96.8} & 94.32 & \textbf{93.63} \\
\hline
\end{tabular}
}
\end{table}

\begin{figure}[t]
  \centering
  \includegraphics[width=\linewidth]{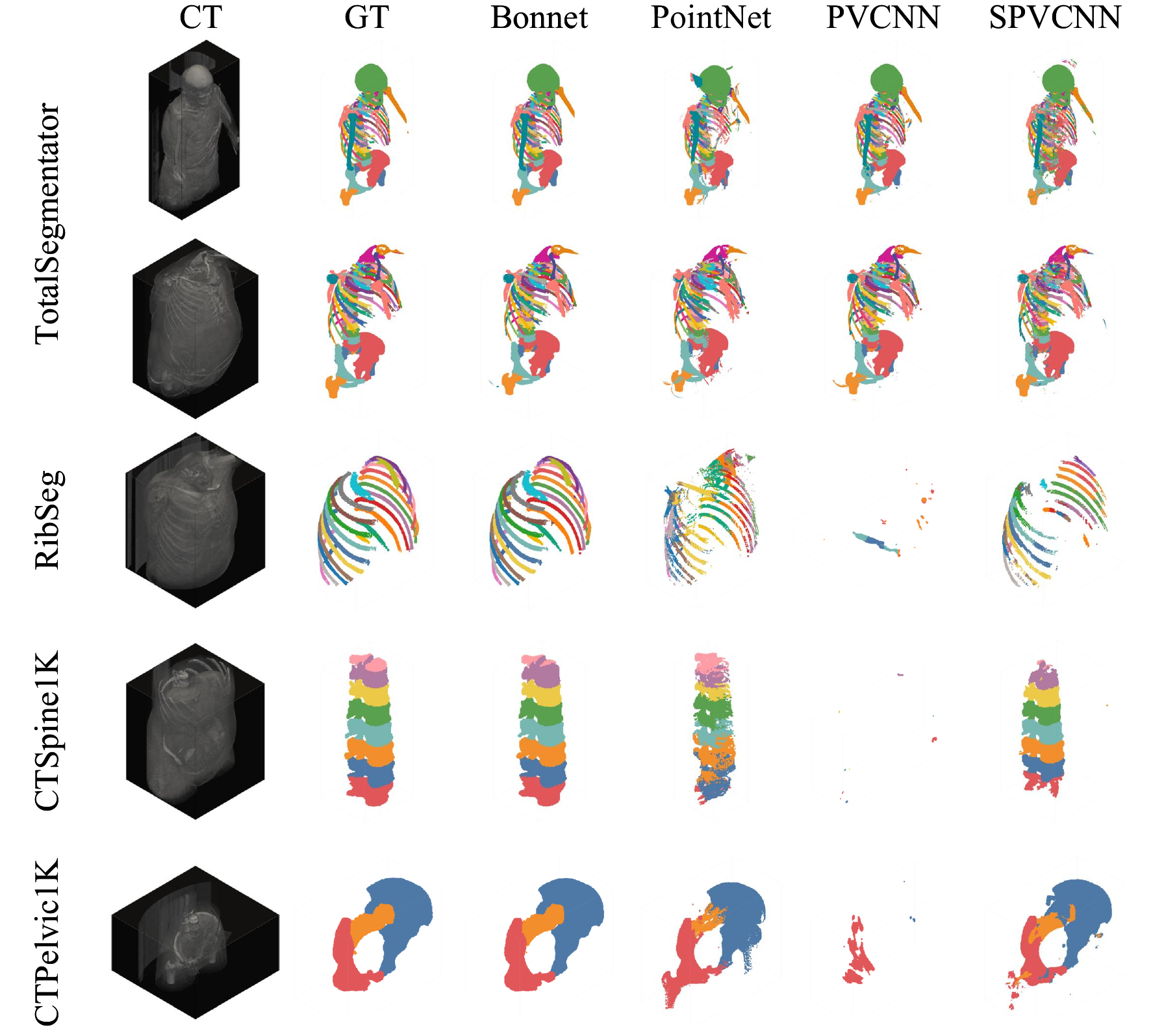}
\caption{Cross-dataset qualitative comparison. From left to right: input CT, ground-truth (GT), Bonnet (ours), PointNet, PVCNN, and SPVCNN. Each row shows an example from the in-domain TotalSegmentator test set and from three external datasets (RibSeg, CT\text{-}Spine1K, and CT\text{-}Pelvic1K).}
  \label{fig:result}
\end{figure}

\subsection{Baseline Methods}
We compare Bonnet against voxel-based and point/sparse baselines. For voxel-based baselines, we include nnU-Net~\cite{isensee2021nnunet} trained with our preprocessing; nnU-Net (RAW) (same architecture without HU-based masking); STU-Net~\cite{huang2023stunet} Small/Base/Large; and STU-Net-S/B/L-ft. For nnU-Net and STU-Net-S/B/L we use the official released implementations, keep their default training settings, and only disable data augmentation; these models are initialized from scratch in our setup (i.e., no external pretrained weights). By contrast, the ``-ft'' variants are initialized from large-scale pretrained weights released by the authors and then fine-tuned on our task. As point/sparse baselines we include PointNet~\cite{qi2016pointnet}, PVCNN~\cite{liu2019pvcnn}, and SPVCNN~\cite{tang2020searching}. Our method, Bonnet, is trained \emph{from scratch} on the TotalSegmentator training split~\cite{doi:10.1148/ryai.230024} using our HU-thresholded sparse pipeline, without any external pretraining or region-specific fine-tuning. The same Bonnet checkpoint is then applied \emph{without retraining} to external datasets covering different skeletal regions (RibSeg for ribs, CT\text{-}Pelvic1K for the pelvis, and CT\text{-}Spine1K for the spine).

\subsection{Results and Discussion}
Bonnet achieves strong accuracy while substantially reducing latency, demonstrating a $\sim\!25\times$ speedup over voxel baselines. On the TotalSegmentator test set (Table~\ref{tab:indomain}), Bonnet reaches 94.25\% Dice on ribs, 99.64\% on pelvis, 94.32\% on spine, and 94.91\% overall, with an average forward time of only 2.69 seconds per scan. In contrast, strong voxel baselines such as STU-Net-L-ft and nnU-Net, though achieving similar or slightly higher overall Dice (e.g., 95.52\% for STU-Net-L-ft), require on the order of a minute per scan.

Cross-dataset generalization further highlights robustness. Without any additional tuning or retraining, the same Bonnet checkpoint maintains competitive Dice on external datasets targeting distinct skeletal regions: 85.91\% Dice on RibSeg ribs, 96.8\% on CT\text{-}Pelvic1K pelvis, and 93.63\% on CT\text{-}Spine1K spine (Table~\ref{tab:ood}). Voxel U-Net-style models can generalize stably but incur heavy computational cost. Bonnet narrows the accuracy gap to these voxel baselines while keeping inference in 2.69 seconds and without relying on large-scale pretrained weights or dataset-specific fine-tuning.

Fig.~\ref{fig:result} shows a cross-dataset qualitative comparison. Each row is a representative case from the internal TotalSegmentator test set or from one of the external datasets (RibSeg, CT\text{-}Spine1K, CT\text{-}Pelvic1K). Columns show the input CT, ground truth, Bonnet, and three point/sparse baselines (PointNet, PVCNN, SPVCNN). In-domain, all methods capture the main anatomy, but Bonnet produces smooth, contiguous bone surfaces, including thin curved structures such as ribs and vertebrae. Under cross-dataset evaluation, where appearance and coverage differ from training, Bonnet still preserves overall skeletal structure across ribs, pelvis, and spine. This is consistent with Tables~\ref{tab:indomain} and \ref{tab:ood}, where Bonnet maintains stable Dice while keeping inference at the seconds level.

\section{Conclusion}
\label{sec:conclusion}
We presented Bonnet, a sparse-volume pipeline for fast whole-body bone segmentation from CT. Bonnet is trained from scratch on TotalSegmentator and then applied, without any retraining, to external datasets covering ribs, pelvis, and spine (RibSeg, CT\text{-}Pelvic1K, CT\text{-}Spine1K). It delivers high Dice on all regions while running in only 2.69 seconds per scan, achieving accuracy comparable to strong voxel baselines such as nnU-Net and STU-Net but with roughly $25\times$ lower latency. This makes Bonnet practical for large-scale anatomical analysis and time-critical clinical workflows. Future work will explore using fast, reliable bone masks as structural priors for downstream tasks such as organ localization and surgical planning.

\section{Acknowledgements}
J.Y. was supported by the ELLIS Institute Finland and School of Electrical Engineering, Aalto University. This work was supported in part by the Swiss National Science Foundation. It was also supported in part by the National Science Foundation (NSF) and SC EPSCoR Program under award number \#OIA-2242812. This research used in part resources on the Palmetto Cluster at Clemson University under NSF awards MRI 1228312, II NEW 1405767, MRI 1725573, and MRI 2018069. The views expressed in this article do not necessarily represent the views of NSF or the United States government.

\bibliographystyle{IEEEbib}
\bibliography{refs}

@article{Zhou2025PointCloud,
  author  = {Zhou, J. and Salvatori, M. and others},
  title   = {Point-cloud segmentation with in-silico data augmentation for prostate cancer treatment},
  journal = {Med. Phys.},
  year    = {2025}
}

@article{ribsegv2,
  author  = {Jin, Liang and Gu, Shixuan and others},
  title   = {RibSeg v2: A Large-scale Benchmark for Rib Labeling and Anatomical Centerline Extraction},
  journal = {IEEE Trans. Med. Imaging},
  year    = {2023}
}

@inproceedings{ribsegv1,
  author    = {Yang, Jiancheng and Gu, Shixuan and others},
  title     = {RibSeg Dataset and Strong Point Cloud Baselines for Rib Segmentation from CT Scans},
  booktitle = {MICCAI},
  year      = {2021}
}

@inproceedings{yang2019modeling,
  title={Modeling point clouds with self-attention and gumbel subset sampling},
  author={Yang, Jiancheng and Zhang, Qiang and others},
  booktitle={CVPR},
  pages={3323--3332},
  year={2019}
}

@article{s24134076,
  author  = {Zhu, Darong and Wang, Diao and others},
  title   = {Research on Three-Dimensional Reconstruction of Ribs Based on Point Cloud Adaptive Smoothing Denoising},
  journal = {Sensors},
  year    = {2024}
}

@article{XIE2025103367,
  author  = {Xie, Kangxian and Yang, Jiancheng and others},
  title   = {Efficient anatomical labeling of pulmonary tree structures via deep point-graph representation-based implicit fields},
  journal = {Med. Image Anal.},
  year    = {2025}
}

@misc{deng2024ctspine1klargescaledatasetspinal,
  author        = {Deng, Yang and Wang, Ce and others},
  title         = {CTSpine1K: A Large-Scale Dataset for Spinal Vertebrae Segmentation in Computed Tomography},
  year          = {2024},
  archivePrefix = {arXiv}
}

@misc{liu2021deeplearningsegmentpelvic,
  author        = {Liu, Pengbo and Han, Hu and others},
  title         = {Deep Learning to Segment Pelvic Bones: Large-scale CT Datasets and Baseline Models},
  year          = {2021},
  archivePrefix = {arXiv}
}

@article{doi:10.1148/ryai.230024,
  author  = {Wasserthal, Jakob and Breit, Hanns-Christian and others},
  title   = {TotalSegmentator: Robust Segmentation of 104 Anatomic Structures in CT Images},
  journal = {Radiol. Artif. Intell.},
  year    = {2023}
}

@article{minnema2022review,
  author  = {Minnema, J. and Ernst, A. and others},
  title   = {A review on the application of deep learning for CT reconstruction, bone segmentation and surgical planning in oral and maxillofacial surgery},
  journal = {Dentomaxillofac. Radiol.},
  year    = {2022}
}

@article{kuiper2023osteotomy,
  author  = {Kuiper, R. J. A. and Colaris, J. W. and others},
  title   = {Impact of bone and cartilage segmentation from CT and MRI on both-bone forearm osteotomy planning},
  journal = {Int. J. Comput. Assist. Radiol. Surg.},
  year    = {2023}
}

@article{isensee2021nnunet,
  author    = {Isensee, Fabian and Jaeger, Paul F. and others},
  title     = {nnU-Net: a self-configuring method for deep learning-based biomedical image segmentation},
  journal   = {Nat. Methods},
  year      = {2021},
  publisher = {Nature}
}

@misc{huang2023stunet,
  author        = {Huang, Ziyan and Wang, Haoyu and others},
  title         = {STU-Net: Scalable and Transferable Medical Image Segmentation Models Empowered by Large-Scale Supervised Pre-training},
  year          = {2023},
  archivePrefix = {arXiv}
}

@article{qi2016pointnet,
  author  = {Qi, Charles R. and Su, Hao and others},
  title   = {PointNet: Deep Learning on Point Sets for 3D Classification and Segmentation},
  journal = {arXiv},
  year    = {2016}
}

@inproceedings{liu2019pvcnn,
  author    = {Liu, Zhijian and Tang, Haotian and others},
  title     = {Point-Voxel CNN for Efficient 3D Deep Learning},
  booktitle = {NeurIPS},
  year      = {2019}
}

@inproceedings{tang2020searching,
  author    = {Tang, Haotian and Liu, Zhijian and others},
  title     = {Searching Efficient 3D Architectures with Sparse Point-Voxel Convolution},
  booktitle = {ECCV},
  year      = {2020}
}

@article{9018080,
  author  = {Shi, Shaoshuai and Wang, Zhe and others},
  title   = {From Points to Parts: 3D Object Detection From Point Cloud With Part-Aware and Part-Aggregation Network},
  journal = {IEEE TPAMI},
  year    = {2021}
}

@misc{graham2017submanifoldsparseconvolutionalnetworks,
  author        = {Graham, Benjamin and van der Maaten, Laurens and others},
  title         = {Submanifold Sparse Convolutional Networks},
  year          = {2017},
  eprint        = {1706.01307},
  archivePrefix = {arXiv},
  primaryClass  = {cs.NE}
}

@misc{qi2017pointnetdeephierarchicalfeature,
  author        = {Qi, Charles R. and Yi, Li and others},
  title         = {PointNet++: Deep Hierarchical Feature Learning on Point Sets in a Metric Space},
  year          = {2017},
  eprint        = {1706.02413},
  archivePrefix = {arXiv},
  primaryClass  = {cs.CV}
}

@inproceedings{inproceedings,
  author    = {Suzani, Amin and Rasoulian, Abtin and others},
  title     = {Deep learning for automatic localization, identification, and segmentation of vertebral bodies in volumetric MR images},
  booktitle = {Proc. SPIE},
  year      = {2015}
}

@misc{graham20173dsemanticsegmentationsubmanifold,
  author        = {Graham, Benjamin and Engelcke, Martin and others},
  title         = {3D Semantic Segmentation with Submanifold Sparse Convolutional Networks},
  year          = {2017},
  eprint        = {1711.10275},
  archivePrefix = {arXiv},
  primaryClass  = {cs.CV}
}

\end{document}